\documentclass[prl,showpacs,twocolumn,floatfix]{revtex4}
\usepackage{graphicx}
\begin{document}
\def\rhov{{\mbox{\boldmath{$\rho$}}}}
\def\tauv{{\mbox{\boldmath{$\tau$}}}}
\def\Lambdav{{\mbox{\boldmath{$\Lambda$}}}}
\def\sigmav{{\mbox{\boldmath{$\sigma$}}}}
\def\xiv{{\mbox{\boldmath{$\xi$}}}}
\def\chiv{{\mbox{\boldmath{$\chi$}}}}
\def\oh{{\scriptsize 1 \over \scriptsize 2}}
\def\ot{{\scriptsize 1 \over \scriptsize 3}}
\def\of{{\scriptsize 1 \over \scriptsize 4}}
\def\tf{{\scriptsize 3 \over \scriptsize 4}}
\title{Possible Orientationally Ordered States of Bucky-Cubane}

\author{A. B. Harris}

\affiliation{Department of Physics and Astronomy, University
of Pennsylvania, Philadelphia, PA 19104}
\date{\today}

\begin{abstract}
For temperatures above $T^*=130$K C$_{60} \cdot$C$_8$H$_8$ 
forms a cubic crystal consisting of two interpenetrating
fcc sublattices, one of freely rotating Buckys C$_{60}$ and
the other of orientationally ordered cubane C$_8$H$_8$.
The crystal structure below a discontinuous transition
is found to be orthorhombic, but the nature of the ordering
of the Buckys has not yet been determined.  Here possible
orderings of the Buckys consistent with the size and
symmetry of the orthorhombic unit cell are analyzed.
Most likely inversion symmetry is preserved at the transition,
in which case the small number of possible orderings are
described.  If inversion symmetry is removed, the point group
can be C$_{2v}$ which supports ferroelectricity or 
D$_2$ which can be confirmed by proton NMR measurements.
\end{abstract}
\maketitle

\section{Introduction}
Orientational ordering of molecules has been studied for many
years.\cite{REV1,REV2}  For the present work the
most relevant studies concern the orientational
ordering of the fullerenes C$_{60}$\cite{PAH} into a
Pa$\overline 3$ space group,\cite{RSABH,DAVID,ABHRS}
and the analogous ordering in solid C$_{70}$.\cite{C70,C70B}
More recently cubane (C$_8$H$_8$) has been crystallized and
its properties studied.\cite{CUBE1,CUBE2} Now\cite{PEKKER,BORTEL,NEMES}
the structure I will call Bucky-cubane
has been studied.  This system forms a cubic crystal consisting
of two interpenetrating fcc lattice, one consisting
entirely of Bucky (C$_{60}$) molelcules and the other entirely
of cubane molecules.  For temperature in the range $130 < T < 470$K
the cubane molecules are ordered with their four-fold axes parallel
to the cubic axes of the crystal, whereas the Bucky molecules are
orientationally disordered.\cite{PEKKER,BORTEL,NEMES} Below $T^*=130$K the
Buckys apparently orientationally order in an as yet undetermined
pattern, but the unit cell (see Fig. 1a) is orthorhombic (o-) with
${\bf a}$ and ${\bf b}$ lattice constants which are nearly equal
to ${\bf a}_c/\sqrt 2$  and ${\bf c}$ nearly equal to
$2{\bf a}_c$, where ${\bf a}_c$ is the cubic lattice constant
just above $T^*$.\cite{NEMES}  As noted in Ref. \onlinecite{NEMES}
this data makes it clear that the o- unit cell contains
four molecules of each species. As in the case of pure C$_{60}$,\cite{ABHRS}
the possible ordered structures which are consistent with this
information can be severly restricted.  

\begin{figure} [ht]
\begin{center}
\includegraphics[width=3.8 cm]{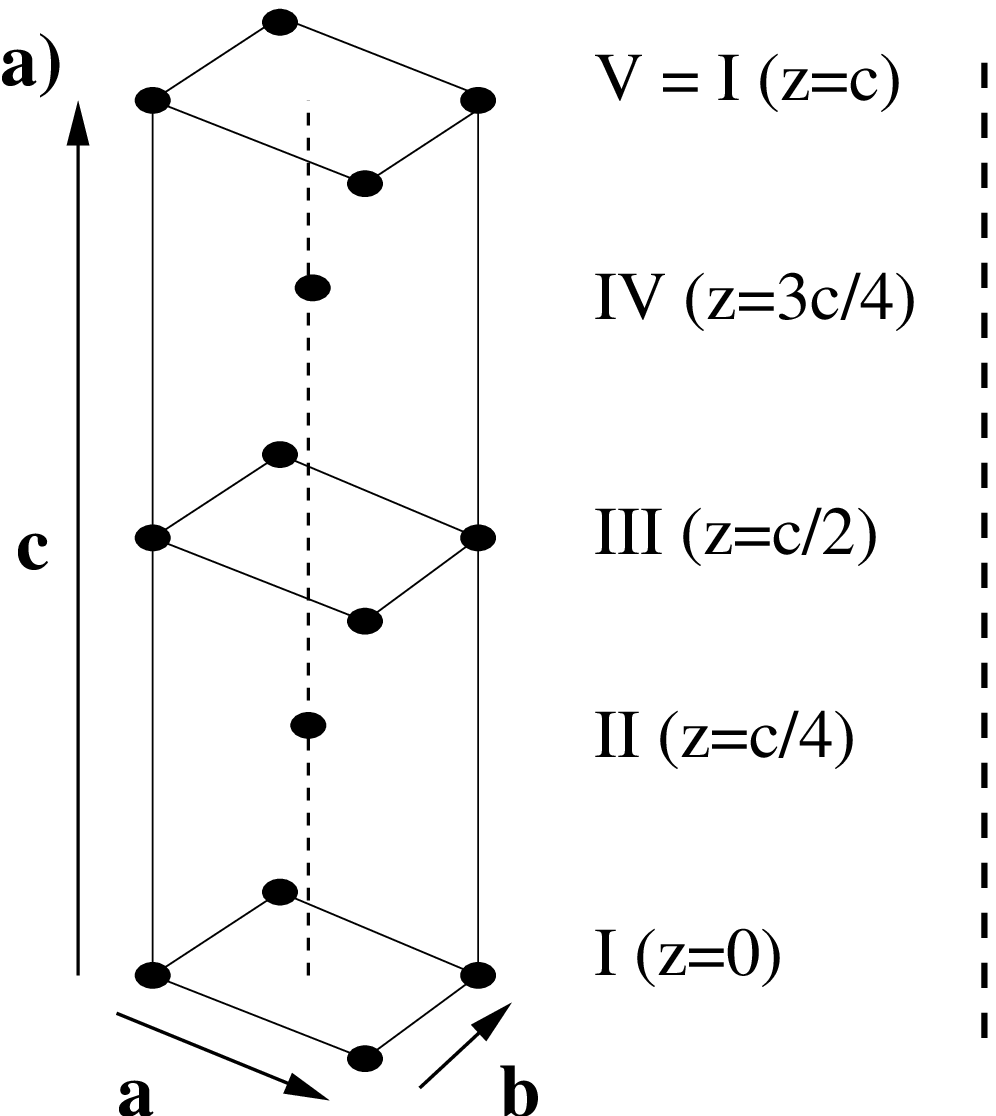}
\hspace{0.18 in}
\includegraphics[width=3.8 cm]{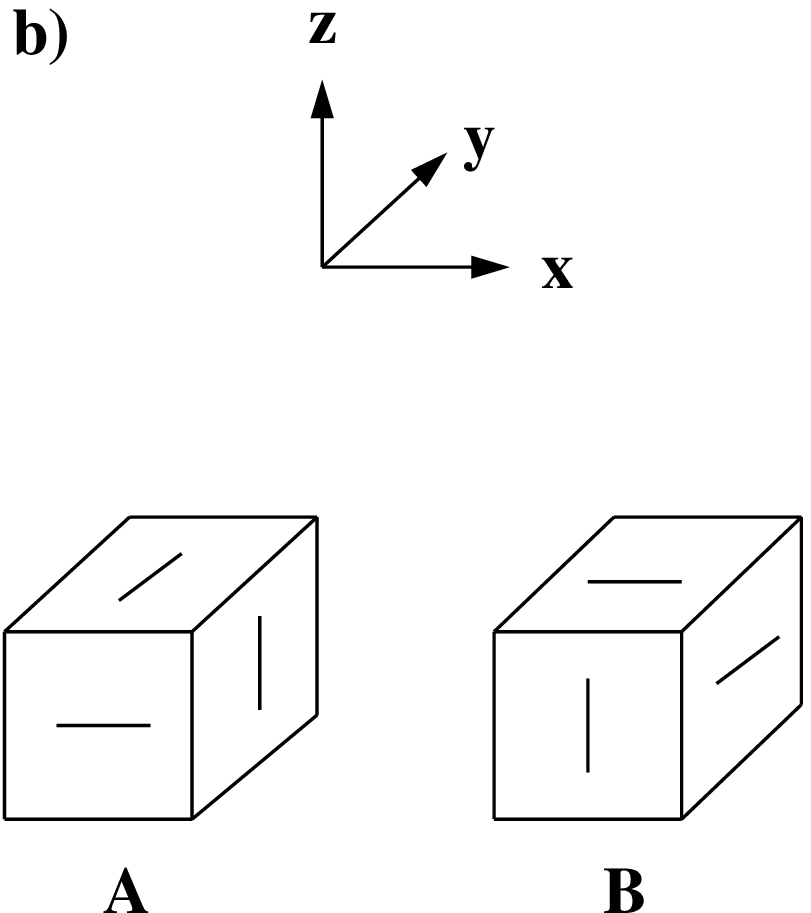}
\end{center}
\caption{a) The o- unit cell, showing only the Buckys, which form
an fcc lattice when the Buckys rotate freely in which case
$a=b=\sqrt 2 c/4$ with $a_c=\sqrt 2 a$  and the cubanes also form
an fcc lattice.  The Buckys form layers perpendicular to the
$c$-axis labeled I-V, as shown.  b) The two standard orientations of
the icosahedron with 12 vertices within a circumscribed cube
relative to the o- axes $x$, $y$, and $z$.\cite{FN1}
Only edges of the icosahedron within the
faces of the cube are shown.  The other edges of the icosahedron are
formed by joining adjacent ends of the edges which lie in the cube
faces. C$_{60}$ is an inversion symmetric truncated icosahedron,
formed by cutting off each of the twelve vertices leaving a
pentagonal face to yield a structure with 60 vertices.}
\end{figure}

In this paper I analyze such possible orientationally ordered
structures under three fundamental assumptions, namely 
1) the indexing to an o- unit cell having the volume
appropriate to four Buckys per unit cell is correct,
2) the cubane molecules retain the static orientation they had
for $T>T^*$ and 3) the locations of the centers of all the
molecules are only perturbatively deformed from their positions for
$T>T^*$. There are three o- point groups,
D$_2$ (222), C$_{2v}$ (mm2), and D$_{2h}$ (mmm). If the
orientational ordering transition were continuous, one could assert
that only a single symmetry is broken at the transition, since
breaking more than one symmetry would require the accidental
degeneracy implied by a multicritical point.  This reasoning
does not apply, of course, to a discontinuous transition as
one has here. However, I regard it is less likely to break
more than one symmetry at a time and therefore I deem it less
likely, but still possible,
that the point group of the ordered phase is D$_2$ or
C$_{2v}$, both of which lack inversion symmetry.  In principle,
the point group can be determined by macroscopic measurements.
For instance, only the point group C$_{2v}$ permits a spontaneous
polarization. In the unlikely event this is the point group of
Bucky-cubane for $T<T^*$, one would have the fascinating appearance
of ferroelectricity driven by molecular ordering.  Otherwise,
in principle, one can distinguish between the other two point group
by NMR experiments, since in D$_2$ the unit cell contains eight
inequivalent proton sites, whereas in D$_{2h}$ there are only
four inequivalent proton sites. (The protons are at nonspecial
positions, so the number of inequivalent protons is equal to
the total number of protons in the unit cell divided by the
number of space group elements.)

In the remainder of this paper, I will concentrate on the scenario
I deem the most likely,  namely that the point group for
$T<T^*$ is D$_{2h}$. In that case, I find that there are four
possible space groups. In one of these the allowed orientations
of the Buckys are quantized, whereas in the other three, the
orientations of the Buckys are characterized by a rotation
angle $\phi$ relative to a ``standard" orientation (see Fig. 1b)
of the C$_{60}$ molecule. Identifying these structures drastically
reduces the number of fitting parameters and should facilitate
a definitive structural analysis from scattering data.\cite{PAH2}

This paper is organized as follows.  First I make some
remarks on the nature of orientational ordering.  Then I
analyze, for the most likely point group D$_{2h}$, the
possible structures within the four allowed space groups.

\section{General Remarks}

As noted in Ref. [\onlinecite{NEMES}] the appearance of a structure
in which the cubanes are orientationally ordered but the Buckys are
orientationally disordered is explained by the orientational
potential on the Buckys being dominated by the cubane-Bucky
interaction.  However, they also say that ``The effect of interfullerene
interactions is negligible, as such interactions would prefer the
formation of the simple cubic Pa$\overline 3$ space group..."
I suggest that the Bucky ordering does result from Bucky-Bucky
interactions in combination with the larger single molecule terms,
as is reflected in the structure of the generalized
orientational Landau free energy
\begin{eqnarray}
F &=& \sum_{M,N,{\bf q}} [\chi(M,N,{\bf q})]^{-1}
Q_6^M({\bf q}) Q_6^N(-{\bf q}) + {\cal O}(Q^3) \ ,
\end{eqnarray}
where $Q_6^M({\bf q})$ is the spatial Fourier transform of
the order parameter $Q_6^M({\bf q})$, a sixth rank tensor which
characterizes, at lowest tensor order, the orientational
fluctuations of the Buckys\cite{ABHRS} and $[\chi(M,N,{\bf q})]^{-1}$
is the inverse orientational susceptibility.  Here
I do not include terms involving products of three or more order
parameters or higher order tensor order parameters.  The first point
is that if one has a sharp orientational ordering transition,
this can not be caused solely by a single molecule potential on the
Buckys due to their interactions with the cubanes.  Although the
orientational transition is not continuous, I nevertheless
expect that the instability toward ordering is signalled by the
minimum in $[\chi]^{-1}$.  While it is true that this minimum
leads to Pa$\overline 3$ ordering in pure C$_{60}$, here the
single molecule terms can shift the minimum to one
corresponding to the actual ordering in Bucky-cubane.
The Bucky-Bucky interactions must
determine the long-range order, as otherwise structures in
which an entire plane of Buckys are in the A standard orientation
would not be favored
over some a distribution of A and B orientations.  Here
I will not attempt any calculations of the
orientational susceptibility, $\chi(M,N,{\bf q})$.

The formal theory outlined above can, in principle, lead
to a prediction of the orientationally ordered structure.
In addition, one can also consider coupling of the
orientational order parameters to various elastic degrees of
freedom, such as strains and zero wave vector optical phonons.
Consider, for instance, the orientational strain interaction,
$V_{\rm OS}$ of the form
\begin{eqnarray}
V_{\rm OS} &\sim& {\rm const} |Q|^2 \epsilon \ ,
\end{eqnarray}
where $\epsilon$ is a component of the strain tensor. As
soon as ordering in $Q$ develops, this term induces a strain
$\epsilon$ of order $|Q|^2$.
Rather than develop the details of such a theory, it is far
simpler simply to look at the symmetry of the orientational ordering
to infer the components of strain it must induce.  This
approach is adopted below.  Similarly, I will deduce the
displacements within the unit cell which are induced by
orientational ordering.

\section{Allowed D$_{2h}$ Structures}

The approach I adopt is as follows. I initially assume that the positions
of the centers of all molecules are the same in the o-
phase as in the cubic phase, the cubanes assume
the same orientations as in the cubic phase, and that the Buckys
assume fixed orientations.  Only certain
o- space groups permit this.  For each allowable
space group I determine the possible orientations of the Buckys
consistent with the space group symmetry.  Then since the unit
cell no longer has cubic symmetry, it must distort into the
actual o- structure.  Also, the molecules would
suffer very small (possibly undetectable) distortions because
their high molecular symmetry is broken by the much lower
symmetry of the sites they occupy.  Finally, the positions
of the centers of the molecules within the unit cell may be
less restricted in the o- structure than in the cubic
structure.  Of course, I will not estimate these distortions,
but, in the case of the positions of the molecular centers,
will only note their symmetry.

We wish to determine which space groups can accomodate
sites at ${\bf r}_1=(0,0,0)$, ${\bf r}_2=(1/2,1/2,1/4)$,
${\bf r}_3=(0,0,1/2)$, and ${\bf r}_4=(1/2,1/2,3/4)$
in terms of fraction of the the lengths of the o- unit cell,
shown in Fig. 1a.  These coordinates must be interpreted
as being coordinates relative to an arbitrary origin which
need not be the same as that used in the listing in Ref.
\onlinecite{ITC}.  I found it convenient to make this
identification in two stages.  In the first stage I looked
for Wyckoff orbits (generated by the space groups elements)
of four sites in the unit cell which reproduce the above ${\bf r}_n$.
Many such orbits can immediately be rejected.  For instance, 
orbits confined to a single plane obviously are rejected.
Also, one of the coordinates (not necessarily $z$)
should assume values close to $z_0$, $z_0+1/4$, $z_0+1/2$ and $z_0+3/4$
in the orbit.  The only space group having a single orbit which
can reproduce the ${\bf r}_n$ is Pccn (\#56) which gives
the orbit
\begin{eqnarray}
(\tf, \tf,-z) \ \ \
(\of,\of,z) \ \ \
(\tf,\tf,\oh-z) \ \ \
(\of,\of,\oh+z) \ .
\label{56}
\end{eqnarray}
If one sets $z=1/8$ and adds $(-3/4,-3/4,1/8)$ to the
above sites, then one gets the desired ${\bf r}_n$.

We can also reproduce the desired ${\bf r}_n$ using orbits
of one or two sites.  For instance, in Pmmm (\#47)
we have the sites (each orbit is enclosed by a square bracket)
\begin{eqnarray}
\big[ (0,0,0) \big]  \ \ \ \big[ (0,0,\oh ) \big] \ \ \
\big[ (\oh ,\oh ,z) \ \ \  (\oh ,\oh ,1-z) \big]
\label{47}
\end{eqnarray}
where $z=1/4$ gives the desired set of sites.  Also, in Pccm (\#49)
we have the sites
\begin{eqnarray}
\big[ (0,0,0) \ \ \ (0,0,1/2) \big] \ \ \
\big[ (\oh ,\of ,\of ) \ \ \ (\oh ,\oh , \tf ) \big] 
\label{49}
\end{eqnarray}
and in Pcmm (equivalent to Pmma, \#51) we have the sites
\begin{eqnarray}
\big[ (0,0,0) \ \ \ (0,0,1/2) \big] \ \ \
\big[ (x,\oh ,\of ) \ \ \ (-x,\oh , \tf ) \big] \ ,
\label{51}
\end{eqnarray}
where $x=1/2$ gives the desired set of sites.

It remains to determine the allowed orientations of the
Buckys on the sites of each of the four allowed space groups.
To do that I show, in Fig. 1b, the ``standard" orientations of
a Bucky to be those for which each o- direction coincides with
a two-fold axis of the Bucky.  (The Bucky has other two-fold
axes, but those are in nonspecial directions relative to the
o- axes.)  Also, in Table I are listed the displacement (glide)
vectors associated with each mirror $m_\alpha$ (which
reverses the sign of the $\alpha$ coordinate) for the
space groups in question.  Note: the glide vectors depend
on the choice of origin and our choice in Eqs. (\ref{56}-\ref{51})
is the same as that of Ref. \onlinecite{ITC}.

\begin{table}
\caption{Allowed space groups (numbered as in Ref. \onlinecite{ITC})
for the inversion symmetric o- point group $mmm$.
The $r_\alpha$ are the displacment (glide) vectors 
for the mirror $m_\alpha$.}

\begin{tabular} {||c | c c c ||}
\hline \hline
S. G. \# & ${\bf r}_x$ & ${\bf r}_y$ & ${\bf r}_z$ \\
\hline
Pmmm (47) & (0,0,0)       & (0,0,0)       & (0,0,0)        \\
Pccm (49) & (0,0,1/2)     & (0,0,1/2)     & (0,0,0)        \\
Pcmm (51) & (0,0,1/2) & (0,0,0)       & (0,0,1/2)      \\
Pccn (56) & (1/2,0,1/2)   & (0,1/2,1/2)   & (1/2,1/2,0)    \\
\hline \hline
\end{tabular}
\end{table}

\subsection{Space Group \#47, Pmmm}

First consider space group  \#47.  The Bucky sites in layers I and III
have mirror planes perpendicular to the o- axes.
So the edges of the Bucky cubes of Fig. 1b coincide with the o- axes.
Thus all Buckys in layer I must either be standard orientation A
or orientation B and similarly for sites in layer III.
The Bucky sites in layers II and IV with $z=1/4+\delta$ for small
$\delta$ have two mirrors and are interrelated by the
remaining mirror.
The presence of these mirrors force all these molecules to
be in the same standard orientation.  So the five layers of Buckys
perpendicular to the $c$ axis (labeled I-V in Fig. 1a) can either be
\begin{eqnarray}
S_1 &=& A,\ A,\ A,\ A,\ A \ \ \ \ \ {\rm or} \ \ \ \ \
S_2 = A,\ A,\ B,\ A,\ A \
\nonumber \\ && \ \ {\rm or} \ \ \ \ S_3=  A,\ B,\ A,\ B,\ A \ .
\end{eqnarray}
(The roles of A and B can be interchanged within the same structure.)
In structure $S_1$, $\delta=0$ is not allowed, 
lest one have a unit cell of side $c/2$.  This same
requirement holds for the structure $S_3$.  In all cases here
and below, the $a$ and $b$ directions are inequivalent because
of the Bucky orientations as is apparent from Fig. 2.

\begin{figure} [ht]
\begin{center}
\includegraphics[width=5 cm]{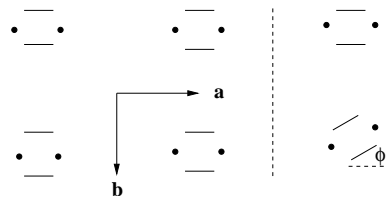}
\end{center}
\caption{Left: A plane perpendicular to the $c$-axis with
all Buckys in the same standard orientation. Only the edges on the
faces of the circumscribed cube of Fig. 1b
not perpendicular to the $c$-axis 
are shown.  The dots represent edges parallel to the $c$-axis.
Clearly, the $a$ and $b$ directions are no longer equivalent in the
presence of the orientationally ordered Buckys. Right: the lower
Bucky is rotated through an angle $\phi$ relative to the upper
Bucky which is in a standard orientation.}
\end{figure}

We make some further comments about the structure $S_n$.  Structure
$S_1$ requires simultaneously breaking orientational symmetry
and enlarging the primitive unit cell.  If the orientational
transition were continuous, this would require a multicritical point,
which one would reject because one never allows an accidental degeneracy
of instability.  But all bets are off for a first order transition.
The same comments apply to structure $S_3$.  Structure $S_2$
will automatically have a distortion ($\delta \not= 0$) because layer II is
not a mirror plane.  So structure $S_2$ is the only one which does not
require the simultaneous breaking of two symmetries and is therefore
preferred on that ground.  However, the o- structure
has very different values for the lattice constants:
$a=9.99$\AA and $b=10.82$\AA.\cite{NEMES}
Structure $S_2$ has three A layers which cause one sign of the
strain making $a-b$ nonzero and a single B layer which
favors the other sign of $a-b$.  In contrast, structure $S_1$
has four A layers and therefore would seem more likely
in view of the large value of $a-b$.  Similarly structure $S_3$
is very bad from that point of view. However, in principle
all three structures are consistent with existing information.
Also note that the Bucky and cubane molecules are
distorted when embedded in the crystal so as to be consistent
with the lower symmetry of the sites they occupy.

\subsection{Space Group \#49, Pccm}

Note that all the mirror operations take layer II into layer IV,
so, by considering the product of two mirrors we conclude that
the sites in layers II and IV have three two-fold axes parallel to
the o- axes.  So these Buckys must all be in the same standard
orientation.  Only the mirror $m_z$ takes the Bucky in layer I
into itself.  So, the Bucky in layer I is obtained from
the standard orientation A by a rotation through an
angle $\phi_z$ about the ${\bf c}$ axis. This orientation is
$A(\phi_z)$, where
$A(\phi_\alpha)$ denotes an orientation obtained from standard
orientation A by a rotation through an angle $\phi_\alpha$
about the $\alpha$-axis.  The mirrors about $x$ and $y$
which take layer I into layer III imply that layer III has
orientation $A(-\phi_z)$.  So one has
\begin{eqnarray}
S_4 &=& A(\phi_z) ,\ A,\ A(-\phi_z) ,\ A,\ A (\phi_z)
\nonumber \\ && {\rm or} \ \ \ \ S_4^\prime =  A(\phi_z) ,\ B,\ A(-\phi_z),
\ B,\ A(\phi_z) \ .
\end{eqnarray}
Structures of type $S_4$ and $S_4^\prime$ are equivalent under
a four-fold rotation about the ${\bf c}$ axis and a relabeling of axes.
Here $\phi_z$ can not assume the special values $\phi_z =0$ or $\pi/2$,
lest the unit cell contain only two Buckys.

\subsection{Space Group \#51, Pcmm}

The sites in layers I and III have the mirror $m_y$ and are related
by the mirrors $m_x$ and $m_z$,
so their orientations are $A(\phi_y)$ and $A(-\phi_y)$,
respectively.  The site symmetry of the sites in layers II and IV
comprises mirror planes $m_y$ and $m_z$ and are interrelated by
$m_x$.  Therefore these Buckys are all in the same standard orientation.
However, the Buckys in layer II are at $x=1/2+\epsilon$ and those in
layer IV are at $x=1/2-\epsilon$, where $\epsilon$ is small.
So the orientations within the layers are
\begin{eqnarray}
S_5 &=& A(\phi_y) ,\ A,\ A(-\phi_y) ,\ A,\ A (\phi_y)
\nonumber \\ && {\rm or} \ \ \ \ S_5^\prime =  A(\phi_y) ,\ B,\ A(-\phi_y),
\ B,\ A(\phi_y) \ .
\end{eqnarray}
Structures of type $S_5$ and $S_5^\prime$ are equivalent under
a four-fold rotation about the ${\bf c}$ axis and a relabeling of axes.
Here $\phi_y=\epsilon=0$ is not allowed lest the unit cell contain only
two Buckys.

\subsection{Space Group \#56, Pccn}

We consider the sites of Eq. (3) with $z=1/8+\eta$ for small $\eta$.
This space group has a two-fold rotation
$m_xm_y$ about (1/4,1/4,0), so the orientation of layer I is
$A(\phi_z)$.  The mirror $m_z$ takes layer I into layer II
and $m_x$ (or $m_y$) takes layer I into layer III and layer II
into layer IV.  These operations fix
the orientation of the other layers in order
of increasing $z$ to be
\begin{eqnarray}
A(\phi_z) \ \ \ A(\phi_z) \ \ \ A(-\phi_z) \ \ \ A(-\phi_z) \ \ \
A(\phi_z) \ .
\end{eqnarray}
Note that $\eta \not= 0$ because no symmetry forces $\eta$ to vanish.

\section{Summary}

To summarize: point groups D$_2$ and C$_{2v}$ break inversion symmetry
and therefore seem unlikely (but still possible) candidate structures
for the phase in which the Buckys are orientationally ordered.
If the phase is ferroelectric it is C$_{2v}$.  If is not
ferroelectric but NMR shows 8 inequivalent protons in the unit cell,
then the phase is D$_2$.  Otherwise, the orientations of the Buckys
in space groups Pmmm, Pccm, Pcmm, and Pccn  are given in Eqs. (7), (8),
(9), and (10),
respectively and the displacements of the layers are discussed in
connection with these equations.

I thank J. E. Fischer for alerting me to this
system and  N. M. Nemes for helpful correspondence.


\begin{thebibliography} {99}
\bibitem{FN1}
The $a$, $b$, and $c$ axes are often referred to as $x$, $y$, and $z$.
\end{thebibliography}
\end{document}